\documentclass[aps,prl,twocolumn,english,showpacs]{revtex4-1}
\usepackage{graphicx}
\graphicspath{ {./Images/} }
\usepackage{amssymb}
\usepackage{amsmath}
\usepackage{epstopdf}
\usepackage{siunitx}
 \usepackage{enumitem}
 \usepackage[usenames, dvipsnames]{color}
\definecolor{MyRED1}{rgb}{ 0,  0,  0}
\definecolor{MyRED2}{rgb}{ 0,  0,  0}

\usepackage{hyperref}
\hypersetup{
  colorlinks   = true,
  urlcolor     = blue,
  linkcolor    = blue,
  citecolor   = blue
}

\newcommand{\shrteq}{\hspace{-0.3em}=\hspace{-0.25em}}

\begin{document}

\title{Fast non-destructive parallel readout of neutral atom registers in optical potentials}

\author{M. Martinez-Dorantes, W.Alt, J. Gallego, S. Ghosh, L. Ratschbacher, Y. V\"olzke, and D. Meschede}
\affiliation{Institut f\"ur Angewandte Physik der Universit\"at Bonn, Wegelerstrasse 8, 53115 Bonn, Germany}

\begin{abstract}
We demonstrate the parallel and non-destructive readout of the hyperfine state for optically trapped $^{87}$Rb atoms. The scheme is based on state-selective fluorescence imaging and achieves detection fidelities $>98$\% within $10\,$ms, while keeping 99\% of the atoms trapped. For the read-out of dense arrays of neutral atoms in optical lattices, where the fluorescence images of neighboring atoms overlap, we apply a novel image analysis technique using Bayesian inference to determine the internal state of multiple atoms. Our method is scalable to large neutral atom registers relevant for future quantum information processing tasks requiring fast and non-destructive readout and can also be used for the simultaneous read-out of quantum information stored in internal qubit states and in the atoms' positions.
\end{abstract}


\maketitle
Cold neutral atoms trapped in optical lattices offer a versatile platform for operating scalable quantum processors ranging from a few to hundreds of qubits: On the one hand, the identity of all atoms makes the system accessible for large-scale global quantum operations. On the other hand, the development of single site detection~\cite{bakr2009,sherson2010} and addressability~\cite{weitenberg2011single} in so-called \emph{quantum gas microscopes} has opened the route to set and read out every qubit individually. Finally, coherent interactions for operating quantum gates can be induced through controlled atom transport and on-site collisional phase shifts~\cite{mandel2003,anderlini2007controlled}.

The standard procedure to detect a qubit encoded in the atomic hyperfine state, however, hitherto proceeds by pushing atoms in one hyperfine states out of the lattice by strong resonant laser radiation and imaging the remaining atoms. As this ``destructive'' detection on average removes half of the atoms from the lattice, the quantum register has to be re-assembled after every read-out, e.g. from a Bose-Einstein condensate of atoms or via atom sorting~\cite{Robens2017sorting}. 
Fast, non-destructive atomic readout, i.e. state detection on the ms-timescale with the atoms remaining in their original optical trapping potential, has previously been achieved for individual atoms coupled to optical cavities~\cite{gehr2010,bochmann2010,reick2010} and for single atoms using state-selective fluorescence detection in free space~\cite{gibbons2011,fuhrmanek2011}. More recently, the explicit detection of both spin states has been demonstrated by using special lattice configurations such as state-dependent potentials~\cite{robens2017atomic} or superlattices~\cite{boll2016spin} to first map the internal spin to position and subsequently use position readout. Thereby both the internal quantum state and the position of the atoms are determined, as needed e.g. in the paradigms of quantum walks and quantum cellular automata~\cite{robens2017atomic,Meyer1996,karski2009quantum}.

Here we present a method for the fast and non-destructive qubit and position read-out of an entire neutral atom quantum register, which represents a powerful feature for the realization of quantum information processing with neutral atoms. By directly detecting and reusing atoms in their optical potentials this readout scheme removes the need for frequent atom reloading and enables the rapid measurement cycles and resource-efficient feedback for quantum error correction, that are considered important for scalable operation: We first state-selectively scatter a small number of photons, sufficient to detect the state of trapped atoms without atom loss. Afterwards we acquire a high signal-to-noise ratio, state-independent image of the same atoms for high-precision position determination, again without atom loss~\cite{Alberti2016SuperResolution}. Accurate models of the statistical and noise properties of the fluorescence and detection processes combined with precise position information permit Bayesian inference-based high fidelity state detection of multiple atoms even from overlapping atom images. For this purpose, we present a scalable Bayesian image processing method for one- and two-dimensional quantum registers.

\begin{figure}[t]
\centering
    \includegraphics[trim = 0mm 0mm 0mm 0mm, width= \columnwidth]{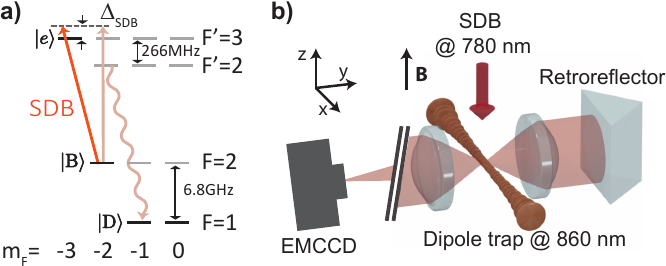}
\caption[Setup] {a) Simplified level scheme of the $^{87}$Rb D2 transition. Only relevant levels are shown.  The $\sigma^-$-polarized \emph{state detection beam}~(SDB) drives the cycling transition $\vert \text{B}\rangle \rightarrow \vert e \rangle$, but polarization impurity can lead to off-resonant scattering via the state $F'=2$ into the dark states $\vert \text{D}\rangle $ (wavy arrow). b)~Simplified experimental setup. Neutral atoms are loaded into a 1D standing-wave optical trap. A six-beam  optical molasses (not shown) is used for \emph{position detection imaging}~(PDI). A single SDB along the $z$ direction is used for state detection imaging. The retro-reflector is used to increase the fraction of fluorescence photons imaged onto the electron-multiplying CCD (EMCCD) camera.} {\label{fig:fig1}
}
\end{figure}

Our state-detection method is based on state-dependent near-resonant fluorescence~\cite{gibbons2011,fuhrmanek2011}: Qubits encoded in the hyperfine ground states of alkali atoms are read out using illumination that resonantly addresses a cycling transition from one ground state while being far detuned from the other ground state. Thus, an atom in the addressed state scatters many photons and becomes \emph{bright}~(B), while atoms in the detuned state remain \emph{dark}~(D). In our experiment we use $\vert \text{B} \rangle \shrteq \vert F \shrteq 2, m_F \shrteq  -2 \rangle $ and $\vert \text{D} \rangle \shrteq \vert F \shrteq 1, m_F \shrteq  0, \pm1,\rangle $ of the $^{87}$Rb atom as Bright and Dark states~\footnote{Whereas we have implemented $m_F$-pumping to initialize the bright state in the experiment, we only use $F$-state pumping for the preparation of the dark states, since all Zeeman sub-levels of the $F=1$ state are equally dark.}. The cycling transition $\vert \text{B}\rangle \rightarrow \vert e \rangle \shrteq  \vert F'\shrteq  3, m_F \shrteq -3\rangle$ is driven by a purely $\sigma^-$-polarized light field to avoid leakage of bright atoms into the dark states via off-resonant excitation of the $\vert F'\shrteq  2 \rangle$ states. This polarization condition requires the use of a single state detection beam (SDB) propagating along the quantization axis (see Fig.~\ref{fig:fig1}a). To this end, we carefully align the SDB to the magnetic bias field and to the electric field vector of the linearly polarized optical dipole trap (see Fig.~\ref{fig:fig1}b).

The drawback of using a single SDB is the absence of laser cooling in all three dimensions. Thus, the average number of photons an atom can scatter before it is lost from the trap, due to unavoidable recoil heating, is $U_0/(2 E_\text{rec})$, where $E_\text{rec}$ is the photon recoil energy and $U_0$ the trap depth. The proportionality of the photon number to the trap depth suggests to use deep optical lattices for scattering a sufficiently large number of photons without losing the atom. In such steep traps, however, another often ignored heating effect plays a detrimental role: \textit{dipole force fluctuations} (DFF) of the trapping potential~\cite{dalibard1985,cheuk2015quantum}: While the optical trapping force is attractive for an atom in the electronic ground state $\vert g \rangle$, it is different, typically repulsive, for the excited state $\vert e \rangle$. Therefore, any change of the internal state induced by photon scattering leads to additional DFF heating. 

When the SDB is weak and on-resonance, the internal atomic dynamics can be described by quantum jumps between ground and excited states. In a very simple model, which assumes a perfectly flat excited state potential $U_e = 0$, we find an exponential DFF heating with rate $\dot E/E = 2 U_{g}^{\prime \prime}/(m\Gamma ^{2})$ ~\footnote{We have used a
harmonic approximation for the ground state potential and assumed that the excited state lifetime $1/\Gamma$ is much shorter than the trap oscillation period}, where $U_{g}^{\prime \prime}$ is the curvature of the trapping potential, $E$ is the total energy, $m$ the mass of the atom, $\Gamma$ the exited-state decay rate and $\dot E$ is the average energy gain~\cite{LongPaper}. This result indicates that for steep optical traps with tight confinement (e.g. in optical lattices with $U_0>\SI{300}{\micro\kelvin} \cdot k_\text{B}$), DFF becomes the dominating source of heating.

Fortunately, DFF can be suppressed by choosing a significant detuning ($|\Delta_\text{SDB}|\gg\Gamma$) of the SDB from the cycling transition: The SDB then dresses the atom, in addition to the dressing by the dipole trap, and scattering of photons happens predominantly without changing the trapping potential. This effect has been recently described for Raman cooling in optical lattices~\cite{cheuk2015quantum}, and an extensive quantitative analysis can be found in Ref.~\cite{LongPaper}.

\begin{figure}[t]
\centering
    \includegraphics[trim = 0mm 0mm 0mm 0mm, width=1\columnwidth]{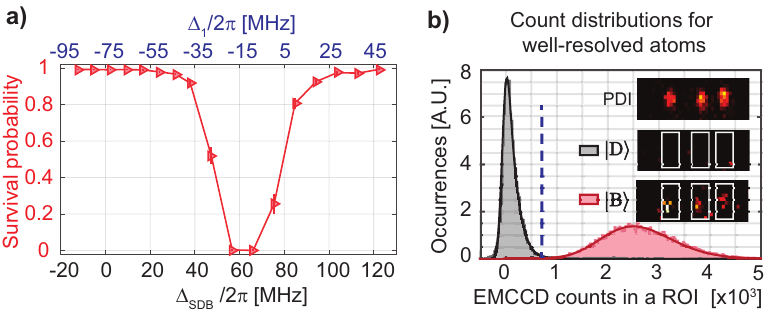}
\caption[Setup] {a)~Measured survival probability of atoms after scattering on average 1100 photons in a deep dipole trap (trap depth $U_0 =  \SI{3.5}{\milli\kelvin} \cdot k_\text{B}$) as a function of the detuning of the state detection beam (SDB). The bottom axis shows the detuning $\Delta_\text{SDB}$ with respect to the bare atomic transition $\vert \text{B}\rangle \rightarrow \vert e \rangle$; the top axis gives the detuning $\Delta_1$ with respect to same transition AC-Stark-shifted by the optical dipole trap. The red solid line is a guide to the eye. b)~Histogram of CCD camera counts for single atoms in the \emph{dark} states $\vert \text{D} \rangle$ (black) and  in the \emph{bright} state $\vert \text{B} \rangle$ (red) using SDB illumination for 10\,ms with a detuning of $\Delta_\text{SDB}/2\pi= +123$\,MHz, and an intensity $I = 1.9 I_\text{sat}$. The solid line is a fit to the camera model (see text). The vertical blue dashed line represents the threshold used to distinguish between the two states for simple threshold analysis. The insets show images obtained using \emph{position detection imaging} (PDI) and state detection imaging of atoms in the $\vert \text{B} \rangle$ and $\vert \text{D} \rangle$ states.
} {\label{fig:fig2}}
\end{figure}

In our experiment, we transfer about 10 laser-cooled $^{87}$Rb atoms from a magneto-optical trap into the standing-wave optical dipole trap (wavelength: 860 nm, beam waist diameter: $\SI{10}{\micro\meter}$). Then, state-independent position detection imaging (PDI) is performed by illuminating the trapped atoms with a six-beam optical molasses (including repumping beam) for \SI{20}{\milli\second} and imaging the fluorescence onto an electron-multiplying CCD (EMCCD) camera, see Fig.~\ref{fig:fig1}(b). As the molasses provides three-dimensional cooling of the atoms, a large number of photons can be scattered, and the resulting high signal-to-noise ratio images are used to determine the number and the position of the atoms in the trap~\cite{Alberti2016SuperResolution}.

To experimentally verify the properties of DFF heating, after PDI the atoms are prepared in state $\vert \text{B} \rangle$ and illuminated with the SDB for different SDB detunings $\Delta_\text{SDB}$, while recording the scattered light with the EMCCD camera. The fraction of surviving atoms is determined from  a second PDI after the SDB illumination. From measurements with different SDB illumination times we determine the fraction of atoms that remain trapped after scattering about 1100 photons, corresponding to about 31 detected photons~\cite{LongPaper}. Fig.~\ref{fig:fig2}(a) shows the resulting probability of bright atoms to remain trapped during SDB illumination: While the survival probability drops to zero close to the AC-Stark shifted resonance due to strong DFF, it remains high for large detunings $\Delta_\text{SDB}$. This shows that, contrary to frequent assumption, resonant illumination is not a good choice for state detection, and large detunings are necessary to suppress DFF. We choose a detuning of $\Delta_\text{SDB}/2\pi= +\SI{123}{\mega\hertz}$ with respect to the free-space cycling transition ($+\SI{44}{\mega\hertz}$ with respect to the AC-Stark shifted atom at the bottom of the optical trap) and an intensity of $I = 1.9 I_\text{sat}$. With these settings, the atoms remain trapped in the same lattice site with a probability of 98.8(2)\%. The detected 31 photons per atom is significantly higher than what has been achieved in other systems~\cite{gibbons2011,fuhrmanek2011}. Moreover, we have measured a very low leakage probability to the dark states of about 2\% despite the fact that frequency selectivity is reduced due to the large detuning of the SDB from the cycling transition.
 
To determine the atomic state from the images obtained of single, optically well resolved atoms, one can simply apply a threshold to the integrated photon counts detected within a certain region of interest around the atom to infer the atomic state. Fig.~\ref{fig:fig2}(b) shows the corresponding count histograms for bright and dark atoms. Using a  threshold discrimination, we obtain a mean detection error of 1.4(2)\%~\cite{LongPaper}. Due to the spatial integration, this simple threshold method however misses the information contained in the spatial distribution the detected photons: Pixels far from the atom's position carry less information on the atomic state than closer pixels. To weight the pixels properly, we make use of Bayesian inference~\cite{sivia2006data}. This allows us not only to achieve higher state detection fidelities, but also to the determine the state of multiple atoms with overlapping fluorescence images, where integrated count histograms would not be well separated anymore and a threshold state discrimination is thus not applicable.

\begin{figure}[t]
\centering
    \includegraphics[width=1.0\columnwidth]{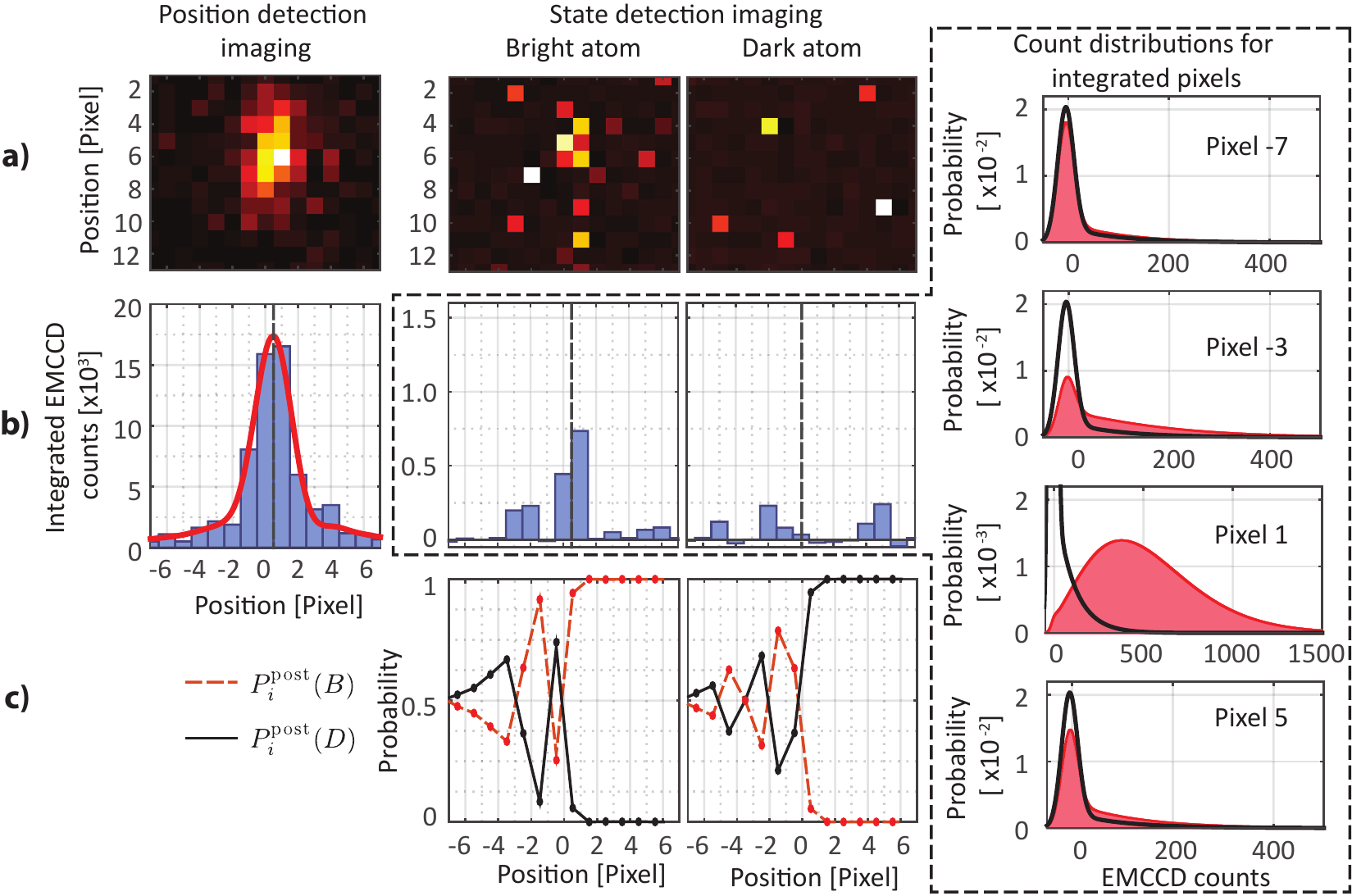}
  	\caption[Setup] {Bayes' method applied to state detection of a well-resolved atom. a)~Position detection imaging of a single atom in the dipole trap (left), state detection imaging of a bright atom (middle) and dark atom (right).  b) Vertically integrated counts for the images above. The vertical dashed line indicates the position of the atom determined from the position detection image.  The plots at the right side show the calculated count distributions $P_{i}(c|\text{S})$ for different pixel columns for a bright (red shading) and dark (black curve) atom. The stronger the bright and dark histogram of a column differ, the more information about the atomic state is contained in this column. c) State determination using the Bayesian update algorithm for a single atom prepared in the bright (left) and dark states (right). See text for details.} {\label{fig:fig3}}
  	\end{figure}

The Bayesian inference relates the measured count distributions for the \emph{bright} and the \emph{dark} states (see Fig.~\ref{fig:fig2}b), which represent the probabilities $P(c|\text{S})$ to detect $c$ counts from an atom in a known state $\text{S}\in\{\text{B,D}\}$, to the desired probability $P(\text{S}|c)$ that an atom is in state S when $c$ counts have been detected: $P\left(\text{S}|c\right)\propto P(c|\text{S})$~\cite{sivia2006data}. The spatial information along the dipole trap axis is included by applying Bayes' theorem to each column $i$ of pixels 
\begin{equation} \label{eq:BayesUpdate}
  P_i^{\text{post}}(\text{S})=P_{i}\left( \text{S}|c\right) =\frac{P_{i}(c|\text{S})P^{\text{pri}}_{i}(\text{S})}{\sum_\text{S} P_{i}(c|\text{S})P^{\text{pri}}_{i}(\text{S})},
\end{equation}
where $P_{i}(c|\text{S})$ is the distribution of counts for column~$i$, and $P_i^{\text{pri}}(S)$ ($P_i^{\text{post}}(S)$) are the probabilities that the atom is in state S before (after) using the information in column $i$. Eq.~(\ref{eq:BayesUpdate}) is applied to each column from left to right, where the result of each iteration is used as a prior for the next one, i.e. $P^{\text{pri}}_{i+1}(\text{S}) = P^{\text{post}}_{i}(\text{S})$ \footnote{using the full two-dimensional intensity distribution, i.e. pixels instead of columns, does not noticeably improve state detection but incurs a drastic increase of computational effort.}.

In contrast to the total count distributions $P(c|\text{S})$, the column count distributions $P_{i}(c|\text{S})$ cannot be measured easily. However, since the spatial distribution of fluorescence photons for bright atoms (i.e. the point spread function), the statistical properties of photon scattering (including the effect of state leakage)~\cite{acton2006}, and the EMCCD camera noise properties are all known~\cite{Robbins2011,citeulike:10216617}, it is possible to accurately calculate the column count distributions (see supplementary material). We illustrate the Bayesian analysis in Fig.~\ref{fig:fig3} for the determination of the state of a well-resolved atom. 

When two atoms are present in the same region of interest, Eq.~(\ref{eq:BayesUpdate}) is used with the combined $2^2=4$ states $\text{S} \in \{\text{BB, BD, DB, DD}\}$. An example for this case is provided in the supplementary material. However, determining the internal state of $N$ atoms this way uses $2^N$ states, for which also the column count distributions have to be calculated. It is thus not scalable to larger quantum registers due to the exponential growth in computational costs.

\begin{figure}[t]
  \includegraphics[width=1\columnwidth]{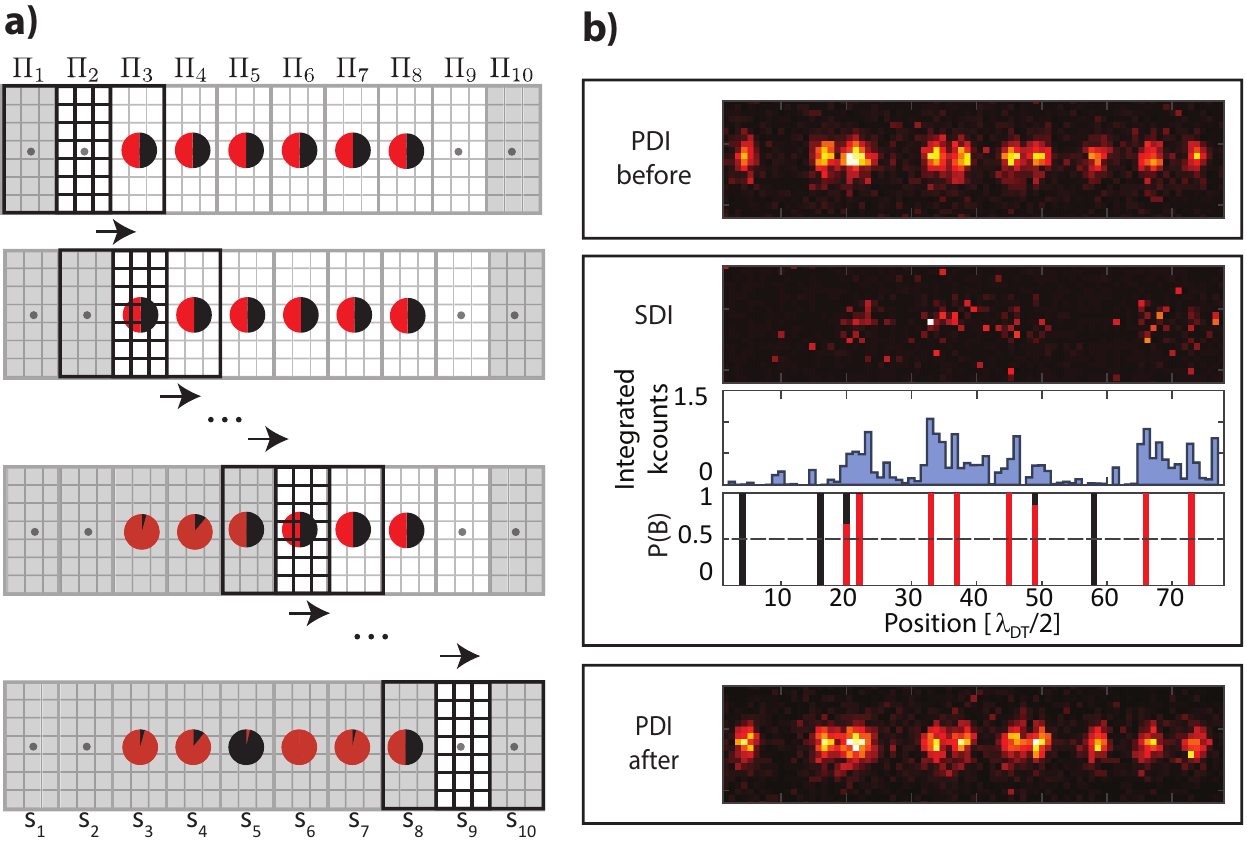}
  \caption{State reconstruction in a one-dimensional \textcolor{MyRED1}{register}. a)~Schematic representation the Bayesian method with shifting patch in one dimension, assuming nearest~neighbor light contamination. The grid represents the pixels of the sensor, the dots represent empty lattice sites, and the circles represent trapped atoms, where the red (black) filling depicts the estimated probability to be bright (dark). A patch is defined around three sites and the middle set of pixels is used to update the combined-state probabilities only for the atoms inside the patch. Then the patch is shifted, the left atom is excluded from the patch by marginalizing its probability, and a new site at the right is included. The shaded regions correspond to pixels that contain either no information or have already been used. b) State dependent imaging (SDI) of a one-dimensional register of neutral atoms (middle box) initialized in random states. The image is integrated along the vertical direction, and the integrated CCD counts are used to calculate the probability that a lattice site contains an atom in the bright state ($P(B)$) using the Bayesian update algorithm. Position detection imaging (PDI) is used before (top box) and after (bottom box) state detection to verify that the atoms remain trapped in the same lattice site.}  {\label{fig:fig4}}
\end{figure}

A more scalable version of this analysis for larger numbers of atoms is obtained by considering that pixels far away from an atom's position do not contain relevant information for this atom. Therefore, instead of applying Bayes' method to all atoms simultaneously, we apply it only to a local image patch containing those atoms whose fluorescence images overlap with that of the central atom of the patch, i.e. those atoms which are informationally linked to the central atom. Bayes' formula is then applied using the central pixels to update the combined state probabilities of the atoms inside the patch. Then the patch is shifted by one lattice site and the procedure is repeated until the information of all pixels has been used and the state probabilities of all atoms have been determined, see Fig.~\ref{fig:fig4}a. A full description of this algorithm is provided in the supplementary material. Fig.~\ref{fig:fig4}b shows an example of the algorithm applied to an image obtained by state detection imaging on a set of atoms where a $\pi/2$ microwave pulse has been used to create a random distribution of bright and dark atoms.

\begin{figure}[t]
		\includegraphics[width=1\columnwidth]{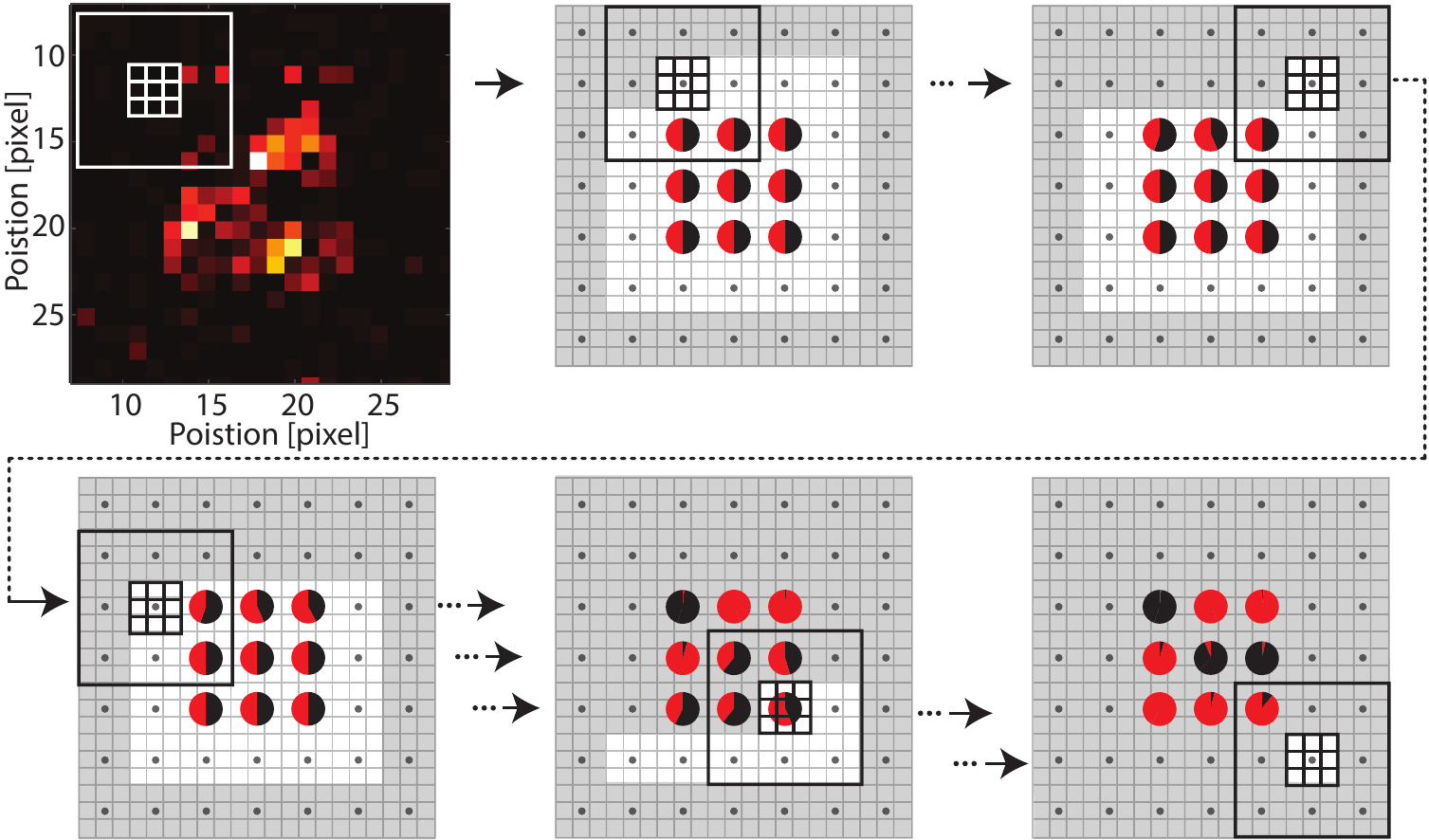}
  	\caption{State reconstruction for atoms in a 2D lattice using Bayes' method. The top left picture is a simulated image. The remaining schematics represent a shifting patch algorithm for the 2D image assuming nearest-neighbor contamination. In each step, the state of the atoms inside the patch is calculated using only the central pixels. The patch is shifted until the last pixel containing information is used.	
  	  	}  {\label{fig:fig5}}
\end{figure}

The same idea is applicable to two-dimensional optical lattices, where typically a significantly larger number of atoms is trapped~\cite{sherson2010,cheuk2015quantum,bakr2009,parsons2015site}. The local patch is shifted row by row over the 2D lattice until the last site has been reached (see Fig.~\ref{fig:fig5}c). Unlike in the one-dimensional case, some of the correlations between state estimates are lost when the patch is moved, and the fidelity slightly decreases. However, we find that this is a good compromise between fidelity and scalability of the algorithm. To benchmark the achievable fidelities, we implement a numerical simulation of atom imaging in a 2D lattice assuming only nearest-neighbor light contamination (see supplementary material). From the simulation we obtain a detection error of 1.4\% using our local patch method. For comparison we also tested other methods commonly employed for atom or state detection in 2D optical lattices: The \emph{threshold method} is faster than our method but has a considerably larger detection error of 8.2\%. The \emph{Lucy-Richardson} deconvolution method~\cite{boll2016spin} requires a similar computation time as our method but has a larger error of 4.8\%. Finally, state determination by fitting multiple point spread functions to the image~\cite{bakr2009} leads to a detection error of 3.3\%, but it is the slowest of all methods.

In conclusion, we have shown that non-destructive high-fidelity state detection of neutral atom quantum registers is possible by combining three major improvements: heating of the atoms due to dipole force fluctuations, leading to rapid trap loss in deep optical lattices, can be reduced by choosing adequate detuning of the state detection beam, allowing us to detect enough photons for spatially resolved state detection; image analysis using Bayesian inference, which properly includes information about the spatial, statistical and noise properties of the experimental setup increases the detection fidelity beyond the fidelities of other methods commonly used on cold atom images; and an adaption of the Bayesian image analysis for multiple atoms provides scalable state detection even for large one- and two-dimensional quantum registers. This fast, non-destructive state detection scheme not only speeds up neutral atom experiments by reusing atoms, but also enables the simultaneous read-out of quantum information contained in the atoms' positions, e.g. in quantum walks~\cite{karski2009quantum}, by following state detection imaging with position detection imaging. In addition, the presented Bayesian image analysis technique presented here is also directly applicable to trapped ion and even solid state quantum systems with imperfectly resolved optical readout. 

We would like to thank Klaus M{\o}lmer and Jean-Michel Raimond for the insightful discussions. This work has been supported by the Bundesministerium f\"ur Forschung und Technologie (BMFT, Verbund Q.com-Q), and by funds of the European Commission training network CCQED and the integrated project SIQS. M.M and J.G. also thank the Bonn-Cologne Graduated School of Physics and Astronomy and L.R. thanks the Alexander von Humboldt Foundation for support.

\newpage

\renewcommand{\thefigure}{S.\arabic{figure}}
\renewcommand{\theequation}{S.\arabic{equation}}
\setcounter{figure}{0}    
\setcounter{equation}{0}    

\section*{Supplementary Material}

\section{Model of the count distribution of the EMCCD camera} \label{sec:EMCCDmodel}

EMCCD cameras are frequently used for applications that require imaging at very low light levels. The spatial resolution provided by EMCCD detectors, however, comes at the price of additional noise contributions compared to Single Photon Counting Modules (SPCMs), such as statistical distribution of the photons over several pixels, clock-induced charges, probabilistic amplification of photoelectrons in the electron multiplication (EM) gain, camera readout noise, dark current, etc. \cite{Robbins2011}.

In our system, we use an EMCCD camera Andor iXon3 DU897D-CS0 to detect the photons scattered by the trapped atoms. The atoms are prepared in the two different states and the measurement --which is described in the main text-- is repeated several times in order to record the detected count distributions for the two states. The resulting camera count distributions for dark and bright atoms are shaped by two main effects: The photon scattering statistics and the camera amplification and noise properties. We first describe the photon scattering process, then derive the camera response, and finally use the combined model to fit the recorded count distributions presented in  Fig.~\ref{fig:fig2}b.

\subsection{Photon scattering statistics}

\subsubsection{Photons detected from a bright atom.}

In an ideal two-level system, the number of photons emitted during the illumination time is Poisson distributed. For a real atom, however, off-resonant excitations can transfer the atom to a dark state, thereby modifying the photon number distribution. This effect is described in Ref.~\cite{acton2006}, which provides an analytic expression for the probability to detect $n$ photons during the illumination process
\begin{eqnarray}\label{eq:AtomPhotonDistribution}
P_{\text{B}}\left( n,\alpha_{\text{B}},n_{0} \right)&=&
\frac{n _{0}^{n}\exp \left[-\left( \alpha_{\text{B}} +1\right) n _{0}\right] }{n!}\\
&&+\frac{\alpha_{\text{B}} }{\left(1+\alpha_{\text{B}} \right) ^{n+1}}\gamma_\text{inc}\left( n+1,\left( 1+\alpha_{\text{B}} \right) n_{0}\right), \nonumber
\end{eqnarray}
where $\gamma_\text{inc}\left( a,x\right) =\frac{1}{\left( a-1\right)}\int_{0}^{x}y^{a-1}e^{-y}\text{d}y$ is the lower incomplete gamma function, $\alpha_{\text{B}}$ is the leakage probability per detected photon from the bright state, and $n_{0}$ is the number of photons that would be detected on average without leakage into the dark state.\\

\subsubsection{Photons detected from a dark atom.}

An atom in the dark state is far detuned and thus scatters only very few photons of the illumination light. However, this off-resonant scattering can transfer the atom to the bright state, where it then scatters a large number of photons. The number of detected photons for an atom initially prepared in the dark state follows the distribution \cite{acton2006}
\begin{eqnarray}\label{eq:AtomPhotonDistributionD}
P_{\text{D}}\left( n,\alpha _{\text{D}}\right)=\exp[-\alpha _{\text{D}}n_{0}]&&\left[ \delta _{n,0}\frac{\alpha _{\text{D}}}{(1-\alpha _{\text{D}})^{n+1}}\right]\\
&&\times \gamma_\text{inc} \left( n+1,\left( 1-\alpha_{\text{D}} \right) n_{0}\right), \nonumber
\end{eqnarray}
where $\alpha _{\text{D}}$ represents the leakage probability from the dark state per detected photon. 

The mean number of detected photons for an atom in the bright state as well as the leakage probability for both states are determined in the following sections using the recorded count distributions. But to this end,  it is necessary to first understand the camera response.

\subsection{EMCCD camera response}

The photoelectrons generated by the detected photons in the EMCCD detector are amplified by electron multiplication (EM) in the ``gain register'', converted into a voltage by the read-out amplifier, and digitized into ``counts'' by the analog-to-digital converter.

\subsubsection{Amplification by electron multiplication and read-out noise}

When $n$ electrons of one pixel are amplified in the gain register, the probability to detect $c$ counts is given by the Erlangen distribution~\cite{citeulike:10216617}
\begin{equation}\label{eq:Erlangen}
P_\text{EM}\left( c,n,\gamma \right) =\frac{1}{\gamma^{n}\Gamma \left( n\right) }c^{n-1}\exp \left( -{c}/{\gamma^{n}}\right),
\end{equation}
where $\gamma$ is the average number of counts after amplification per electron in the pixel and $\Gamma(n)$ is the Gamma function. After the multiplication process, Gaussian noise is added by the read-out amplifier,
\begin{equation}\label{eq:Readout}
P_{\text{read}}\left( c,\sigma,\mu\right) =\frac{1}{\sqrt{2\pi }\sigma}\exp \left[-\frac{\left( c-\mu \right) }{2\sigma ^{2}}\right],
\end{equation}
where $\mu$ is an electronic offset added to the output signal and $\sigma$ is the width of the noise distribution in units of counts. The distribution of counts $c$ after EM amplification and readout for a single pixel containing $n$ electrons is given by the convolution of the probabilities in Eqs.~(\ref{eq:Erlangen}) and (\ref{eq:Readout})
\begin{equation}\label{eq:SinglePixelCounts}
P\left( c,n,\gamma,\sigma,\mu \right) = %
P_\text{EM}\left( c,n,\gamma \right) \ast%
P_{\text{read}}\left( c,\sigma,\mu \right). 
\end{equation}

If $N = \sum_{i=1}^m n_i$ electrons are distributed over $m$ pixels, the probability distribution describing the total number of counts $c$ integrated over the $m$ pixels after the readout is
\begin{eqnarray}\label{eq:NcharegesinMPixels}
P_{m} \left(c,N,\gamma,\sigma,\mu \right)& \equiv & \ast \prod_{i=1}^{m}P\left(c,n_{i},\gamma,\sigma,\mu\right) \\
 & & \mkern-50mu = \!P_\text{EM}\left(c,\!N,\!\gamma \right)\!\ast\! P_{\text{read}}\left( c,\!\sqrt{m}\sigma,\!\mu\right)\!,
\end{eqnarray}
i.e. the read-out noise $\sigma$ is increased by $\sqrt{m}$.\\

\subsubsection{Clock induced charges}

Besides photoelectrons, \emph{clock induced charges} (CIC) are generated randomly by the vertical CCD shift operation. The probability $p_{0}$ to generate a CIC on a pixel is roughly constant throughout the CCD, and hence the number of CIC is Poissonian distributed. In consequence, the probability that $n$ CIC are generated in a set of $m$ pixels is also Poissonian distributed as
\begin{equation}\label{eq:PCIC}
P_{\text{CIC}}\left( n,m\right) =\frac{\left( mp_{0}\right) ^{n}\exp (mp_{0})}{n!}.
\end{equation}

\subsubsection{Total charges on the sensor (photons + CIC)}

The total number of charges generated in the CCD is the sum of the CIC and electrons generated by photon detection. Therefore, the distribution of electrons after the readout process is described by the convolution of the probabilities in Eqs.~(\ref{eq:AtomPhotonDistribution}), (\ref{eq:AtomPhotonDistributionD}), and (\ref{eq:PCIC})
\begin{eqnarray}\label{eq:ProbTotalCharges}
&P_{\text{tot,S}}&\left(n,m,\alpha_{\text{S}},n_{0}\right)= P_{\text{S}}\left( n,\alpha_{\text{S}}, n_{0} \right)\ast
P_{\text{CIC}}\left( n,m\right) \nonumber  \\ 
&=&\sum_{k=0}^{n}P_{\text{S}}\left( n-k,\alpha_{\text{S}} \right) \frac{\left(m\cdot\,p_{0}\right) ^{k}\exp (m\cdot\,p_{0})}{k!}
\end{eqnarray}
for $\text{S}=\text{B},\text{D}$.\\

\subsubsection{EMCCD count distributions for bright and dark atoms}
 
Now, we have all the elements needed to model the count histograms for bright and dark atoms: The probability to generate $N$ charges on the sensor is described by Eq.~(\ref{eq:ProbTotalCharges}), and the camera response to $N$ charges distributed over $m$ pixels is described by Eq.~(\ref{eq:NcharegesinMPixels}). Combining these results we obtain the distributions of EMCCD counts for an atom in the state $\text{S}\in\{\text{B},\text{D}\}$ under illumination 
\begin{eqnarray}\label{eq:DistributionsModel}
D_{\text{S}}(c,n_{0}\ ; \gamma,\sigma, m,\alpha_ {\text{S}})= \ \ \ \ & \nonumber\\
\sum_{N=0}^{\infty }P_{\text{tot,S}}\left(N,m,\alpha_{\text{S}},n_{0}\right)&P_{m}\left(c,N,\gamma,\sigma,\mu\right) 
\end{eqnarray}
Fig.~\ref{fig:fig2}b shows the result for a fit of Eq.~(\ref{eq:DistributionsModel}) to the count histograms for the bright and dark states. From the fit we find that the mean number of detected photons per bright atom is $n_{0} = 31.1$; a probability to generate a CIC of $p_0=0.019$, which also takes into account the contribution from stray light; and a leakage rate per detected photon of $\alpha_\text{B}=0.0010$ and $\alpha_\text{D}=0.0011$, which lead to a total leakage rate of $\sim3\%$ that is in agreement with an independently measured leakage rate for the bright state of $\sim2\%$~\cite{LongPaper}.

The results from the fit can now be used to calculate the count distributions of individual pixel columns from Eq.~(\ref{eq:DistributionsModel}) by setting $m$ equal to the number of pixels per column and using for $n_0$ the average number of photons for a column at a specific distance from the atom, as obtained from a measured point-spread function or line-spread function (LSF) of the imaging system~\cite{Alberti2016SuperResolution}. The count distributions for the pixels columns shown in Fig.~\ref{fig:fig3} have been calculated in this way.

\section{1D Bayes algorithm for two atoms}

In the main text we have used Bayesian inference to determine the state of a single atom. In general, when no prior information is available for Bayes' formula, one can assume a flat distribution for the priors and, in such a case, the Bayesian approach becomes equivalent to the \emph{maximum likelihood estimation} (MLE) method~\cite{sivia2006data}. The MLE has been used, for example, to determine the state of a chain of trapped ions in Ref.~\cite{burrell2010}. In this, section we present the generalization of Bayes' method from a single atom to two atoms.

With two atoms there are four possible outcomes for the readout of the internal states: BB, BD, DB and DD, which represent all possible combinations of bright (B) and dark (D) states. Bayes' formula in Eq.~(\ref{eq:BayesUpdate}) is directly applicable to the two atom case by using $\text{S} \in\{\text{BB, BD, DB, DD}\}$ once all column count distributions $P_{i}(c|\text{BB})$,  $P_{i}(c|\text{BD})$,  $P_{i}(c|\text{DB})$,  $P_{i}(c|\text{DD})$ are determined.

\subsection{Calculation of the column count distributions}

The number of photons detected from the two atoms $k=1,2$ in column $i$ is $n_{k,i} = \operatorname{LSF}(x_k-x_{pi})$, where $x_k$ and $N_k$ are the position and mean number of detected photons for atom $k$, $x_{pi}$ is the position of column $i$, and $\operatorname{LSF}(x)$ is the line spread function normalized in the selected region of interest. The mean number of detected photons is the same for all bright atoms  $N_k=n_0$  and since we detect less than 0.5 photons in average from an atom in the dark state, we assume that $N_k\approx0$ for the dark atoms. 

The number of photons detected from the two atoms in each column is the sum of the photons coming from each atom. Therefore, the distribution of total detected photons in column $i$ is obtained by the convolution of the corresponding distribution for each atom in Eq.~(\ref{eq:AtomPhotonDistribution}).

\begin{eqnarray} \label{eq:TwoAtomPotonDist}
P_{\text{S$_1$,S$_2$}}&(n&,\alpha_{\text{S$_1$}},\alpha_{\text{S$_2$}}n_{1,i},n_{2,i})\nonumber \\
&=& P_{\text{S$_1$}}( n,\alpha_{\text{S$_1$}},n_{1,i}) *P_{\text{S$_2$}}( n,\alpha_{\text{S$_2$}},n_{2,i})
\end{eqnarray}
for  $\text{S$_1$,S$_2$}\in\{ \text{B,D}\} $. Finally, the four count distributions are obtained by replacing $P_{\text{S}}\left( n,\alpha_{\text{S}}, n_{0} \right)$ in Eq.~(\ref{eq:ProbTotalCharges}) with $P_{\text{S$_1$,S$_2$}}(n,\alpha_{\text{S$_1$}},\alpha_{\text{S$_2$}},n_{1,i},n_{2,i})$ and using them in Eq.~(\ref{eq:DistributionsModel}).

\subsection{Experimental state detection of two atoms}

\begin{figure}[t]
\centering
    \includegraphics[width=1.0\columnwidth]{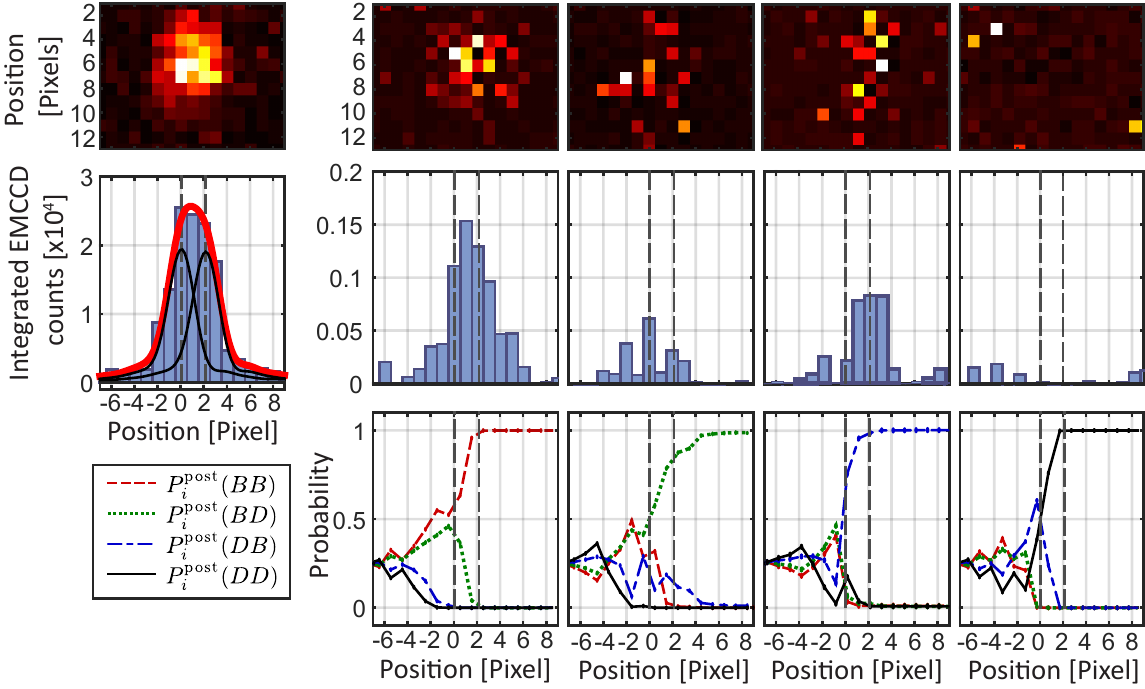}
  	\caption[Setup] {Bayes method applied to the image analysis for two atoms. Top row: Molasses imaging of two atoms separated by two lattice sites in the dipole trap (left) and the state-dependent fluorescence of such atoms prepared in the states BB, BD, DB, and DD.  Middle row: Vertically integrated counts for the reference images (with fit by a sum of two line-spread functions) and for the state detection images. The positions of the atoms are indicated by the vertical dashed lines. Bottom row: State determination using the Bayesian update algorithm for the atoms in all four possible states.} {\label{fig:fig6}}
\end{figure}

In our experimental apparatus, we can prepare atom pairs in either the state BB or DD but we cannot address neighboring atoms individually to create the states BD and DB in a deterministic fashion. Nevertheless, we can use the fact that the ``signal'' from an atom in the dark state is very similar to an empty site and ``simulate'' the dark-state atom by an empty lattice site in the cases BD and DB. Fig.~\ref{fig:fig6} shows an example of the algorithm applied to a pair of atoms merely separated by two lattices sites, where their fluorescence images overlap significantly. 

\begin{figure}[t]
\centering
    \includegraphics[width=1.0\columnwidth]{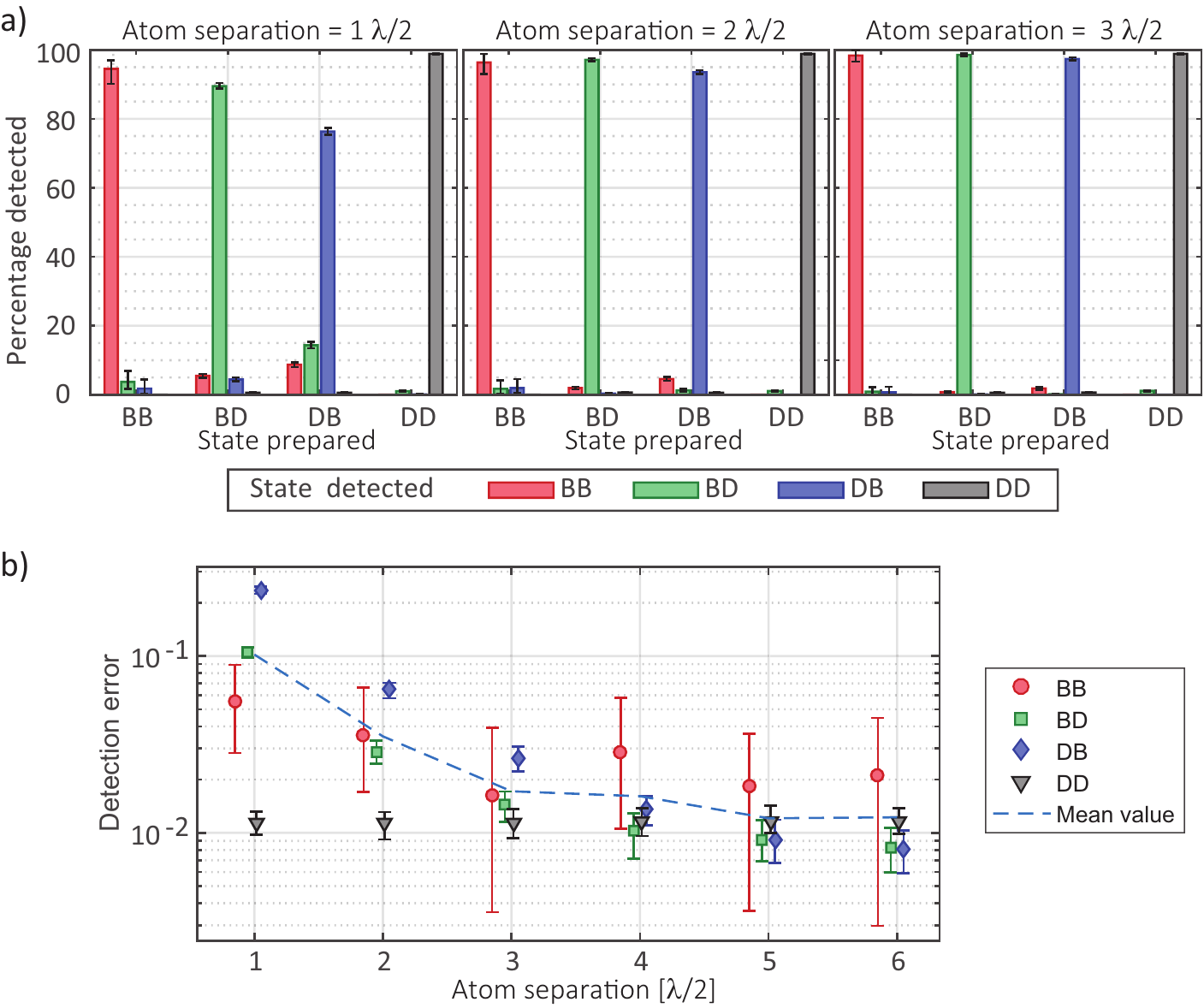}
  	\caption[Setup] {Detection fidelity vs. separation for two atoms.
a)~Detection fidelity for all possible outcomes for atom pairs prepared in different states, $P(\text{S$'$}|\text{S})$ separated by 1, 2 and 3 lattice sites respectively. b) Detection error, $Err(S)$, (markers) and mean error (dashed line) for two atoms separated from one to 6 lattice sites. All error bars represent 95\% confidence interval obtained by bootstrap resampling.} {\label{fig:fig7}}
\end{figure}

To characterize the state detection fidelity, we determine the probability $P(\text{S$'$}|\text{S})$ that an atom pair prepared in a state S is detected in a state S$'$ for ${\text{S,S'}\in\{\text{BB, BD, DB, DD}\}}$. Fig.~\ref{fig:fig7} shows $P(\text{S$'$}|\text{S})$ using images of atoms separated by 1, 2 and 3 lattices sites obtained in our experiment. Even though the state determination for atoms separated by only one lattice site is challenging, the detection of DD and BB states is quite accurate (fidelity $>95\%$), while the states BD and DB are still detected correctly with 85\% probability. One observes that the state BD is detected with higher accuracy than the state DB. This arises from a small asymmetry of our LSF, which creates more light contamination to one side.

For separations of two and three lattice sites the fidelity is high ($>95\%$) for all states, which is quite good taking into account that at these separations the atoms are still not optically resolved. We define the \textit{detection error} for the state S as $Err(\text{S})= 1-P(\text{S}|\text{S})$ and the \emph{mean detection error} as the average value of $Err(\text{S})$ for the four states S. The detection errors are plotted in Fig.~\ref{fig:fig7} for atoms separated by one to six lattice sites.  As expected, the detection error decreases as the distance between the atoms increases and asymptotically approaches the value for the single atom case. 

\section{1D Bayes update algorithm with sliding patch }

For the scalable Bayesian algorithm for long 1D chains of atoms, we assume that a dark atom is indistinguishable from an empty lattice site. 
This avoids complications when including empty lattice sites (where no atom was loaded) into the analysis simply by setting $P^\text{pri}_B(s_{i})=0$ as prior, while treating all lattice sites equally otherwise.

The algorithm is introduced using the example depicted in Fig.~\ref{fig:fig4}a. We consider a region of interest (ROI) containing 10 lattice sites with 6 atoms at the center. It is assumed that the fluorescence of each atom contaminates only directly neighboring sites. The routine is then implemented as follows: 

\begin{enumerate}[leftmargin=0.5cm]

\item The ROI is divided in 10 sets of pixel columns $\Pi_i$ for $i=1,...,10$ (to which we will refer to just as \emph{sets} for this description).  Each set is centered on one lattice site $s_{i}$. Although the first and last set ($\Pi_1$ and $\Pi_{10}$) do not contain any information (photon counts) because their central and their neighboring lattice sites are all empty, these sets are included in order to treat all sites in the same manner. Otherwise, the starting and finishing points would need to be processed in a different way.

\item We define a patch that surrounds the lattice sites $s_{1},s_{2},$ and $s_{3}$ and fully contains the respective sets $\Pi_{1},\Pi_{2},$ and $\Pi_{3}$. Then, Bayes' formula in Eq.~(\ref{eq:BayesUpdate}) is applied only on the middle set ($\Pi_2$) to obtain the probability for the $2^3$ possible combinations $\text{S} = \text{S}_1 \text{S}_2 \text{S}_3$ of bright and dark states of the three sites within the patch. 

\item  The patch is then shifted by removing the left site $s_1$ and including  $s_4$. To remove the site $s_1$ we calculate its bright state probability by marginalization:
\begin{equation}
P_{\text{B}}(s_{1}) =\sum_{\text{S}_{2}\text{S}_{3}\in \{\text{B},\text{D}\}}P^{\text{post}}\left( \text{BS}_{2}\text{S}_{3}\right).
\end{equation}

\item To apply Bayes' update in the new patch, we need to first calculate the new prior probabilities. To this end, we use the result of the calculated posterior probabilities $P^{\text{post}}(\text{S})$  together with the prior for the site that was added $P_{\text{B}}^{\text{pri}}\left(
s_{4}\right)$. The new set of priors are then given by
\begin{equation}
\begin{array}{ccc}
\begin{array}{c}
P^{\text{pri}}\left( \text{S}_{2}\text{S}_{3}\text{B}\right) =P^{\text{post}}\left( \text{BS}_{2}\text{S}_{3}\right) P_{\text{B}}^{\text{pri}}\left(
s_{4}\right)  \\ 
P^{\text{pri}}\left( \text{S}_{2}\text{S}_{3}\text{D}\right) =P^{\text{post}}\left( \text{BS}_{2}\text{S}_{3}\right) P_{\text{D}}^{\text{pri}}\left(
s_{4}\right) 
\end{array}
& \text{if} & P_{\text{B}}\left( s_{1}\right) >0.5  \\
&  &  \\ 
\begin{array}{c}
P^{\text{pri}}\left( \text{S}_{2}\text{S}_{3}\text{B}\right) =P^{\text{post}}\left( \text{DS}_{2}\text{S}_{3}\right) P_{\text{B}}^{\text{pri}}\left(
s_{4}\right)  \\ 
P^{\text{pri}}\left( \text{S}_{2}\text{S}_{3}\text{D}\right) =P^{\text{post}}\left( \text{DS}_{2}\text{S}_{3}\right) P_{\text{D}}^{\text{pri}}\left(
s_{4}\right) 
\end{array} 
& \text{if} & P_{\text{B}}\left( s_{1}\right) <0.5  \\
\end{array}
\end{equation}
for $\text{S}_{2,3}\in\{\text{B,D}\}$. The new priors $P^{\text{pri}}(\text{S})$ are also renormalized. In this way, all the correlations between the state of the atoms that remain inside the patch are maintained.

\item Bayes' formula is applied once more using only the middle set of pixels ($\Pi_{4}$). By doing this, we obtain the posterior probabilities $P^{\text{post}}(\text{S})$ for $\text{S} = \text{S}_2 \text{S}_3 \text{S}_4$. 

\item  The whole procedure is repeated from point 3 until the last set of pixels that contains information is reached. In this example, the last filled lattice site is $s_8$, therefore we continue until we have used set $\Pi_{9}$.
\end{enumerate}

The extension of the algorithm for light contamination larger than one lattice site is straight forward. If the light contaminates $n_c$ lattice sites, it is necessary to include $2n_c+1$ sites in the patch. The application of this algorithm to experimental data is presented in Fig.~\ref{fig:fig4}b. We have considered light contamination of $n_c=3$ lattice sites.

\section{2D Bayes update algorithm with shifting patch}

The extension of the Bayesian shifting patch algorithm for state detection in a two-dimensional system is closely related to the one-dimensional case. Therefore we describe here the example of Fig.~\ref{fig:fig5} with $3\times3$ occupied lattice sites, omitting the somewhat unwieldy equations. Generalization to larger atom arrays is straightforward. In the same way as before, we assume that only nearest neighbor light contamination is present in the image.  

\begin{enumerate}[leftmargin=0.5cm]

\item A region of interest of $7\times7$ lattice sites containing 9 atoms in the central $3\times3$ sites is considered (see Fig.~\ref{fig:fig5}). Note that the pixels corresponding to outermost sites contain no information but are included for consistency of the description.

\item Bayes formula is applied to the first square patch containing $3\times3$ lattice sites (of which 8 are empty), and only the pixels corresponding to the central lattice site are used to calculate the probability for the $2^{9}$ combination of states (where $P^\text{pri}_B(s_{i})=0$ has been set for empty sites). This guarantees that pixels containing information on atoms outside the patch are not used during the update algorithm.

\item The patch is shifted by one lattice site in the horizontal direction. The probability to contain a bright atom for the sites removed from the patch is estimated by marginalization, and the probability values are kept in memory since they will be used at later steps. The correlations between the six atoms remaining inside the patch are maintained and are used as priors together with priors of the new included sites.

\item The patch is shifted to the right until the final lattice site of the row is reached.

\item The patch is shifted one lattice site down and the whole procedure is repeated starting from the beginning of the row. 

\item The algorithm is repeated until the last lattice site is reached.

\end{enumerate}

This algorithm ensures that the information contained in a given pixel is used only once and that the estimated state of all lattice sites that contribute to the signal of a given pixel are updated during its evaluation. However, in contrast to the 1D case, the 2D case requires that lattice sites, the state of which already evaluated by marginalization, are again added to the shifting pattern at a later stage of the shifting process, i.e. not all correlations between sites connected by light contamination can be maintained during the shifting process. The presented shift path is of course not unique and different shifting patterns are conceivable.

\section{Simulation of state detection and comparison with other methods}

In this section, we present details on the numerical simulation used to characterize the fidelity of the Bayesian algorithm methods for state-dependent images of atoms trapped in a 2D lattice. Here the bright atoms are simulated with an average number of 31 photons. The PSF is an ideal Airy function with a Full-Width Half Maximum (FWHM) of three pixels and the lattice spacing is assumed to be one FWHM. The effects of uniform stray light contamination and clock induced charges are included by assuming that each pixel has a probability of 1.9\% to contain a contaminating  photo-electron. The numerical values for the detected photon number and the light contamination are the measured quantities for our system.

The state detection fidelity is quantified by simulating an atom (either dark or bright) surrounded by 8 atoms in different states. The number of bright neighbors is varied from 1 to 8 at random positions.  We define the mean error as the ratio of correctly detected to simulated atoms for all cases.

Bayes' rule is used in two different ways:
\begin{itemize}

\item \textit{Bayesian Method with Global Evaluation} (BMGE). Here, Bayes' method is applied by updating the state of all lattice sites of an array for the evaluation of every pixels.  In the example of Fig.~\ref{fig:fig5} the total number of lattices sites of the array is also only $N=3\times3$ sites and hence $2^9$ combination of states. This method is the most accurate, but the computational effort scales exponentially with the number of lattice sites in the array.

\item  \textit{Bayesian Method with Shifting Patch} (BMSP): This is the method that has been presented in the previous section. Here a patch is defined around nine lattice sites (see Fig.~\ref{fig:fig5}). In each shifting step Bayes' formula is applied only using the central set of pixels to update the relevant $2^9$ states until the last site is reached. 

\end{itemize}

To put the detection fidelity of Bayes' method into context, we compare it to the performance of other known techniques: \textit{Threshold method} (TM). The image is divided in multiple ROI, each one containing a lattice site; depending on the total counts we determine the state of the atom with the threshold that gives the smallest mean error. \textit{Lucie-Richardson deconvolution}(LR). The image is deconvolved using the Lucie-Richardson method, and then the TM is applied to the deconvolved image. \textit{Multiple PSF fit} (MPSF). The image is fitted by the sum of multiple PSFs centered at the lattice sites, and the state of the atom is inferred depending on the fit result of the PSF amplitudes by setting an optimal threshold. We obtain the following mean error for the different methods: TM:~8.2\%, LR:~4.8\%, MPSF: 3.3\%, BMGE:~1.18\% and BMSP:~1.4\%.
The difference on the mean error between the BMGE and BMSP comes from the shifting procedure: When the patch is shifted the state probabilities of the atoms temporarily leaving the pattern have to be obtained by marginalization, and in this process some correlations between the state combination probabilities are lost, leading to a lower fidelity. However, in contrast to BMGE, the BMSP method features a linear dependence of the computational effort/time on the number of atoms and hence is applicable to large arrays.

\bibliography{References}

\begin{thebibliography}{28}%
\makeatletter
\providecommand \@ifxundefined [1]{%
 \@ifx{#1\undefined}
}%
\providecommand \@ifnum [1]{%
 \ifnum #1\expandafter \@firstoftwo
 \else \expandafter \@secondoftwo
 \fi
}%
\providecommand \@ifx [1]{%
 \ifx #1\expandafter \@firstoftwo
 \else \expandafter \@secondoftwo
 \fi
}%
\providecommand \natexlab [1]{#1}%
\providecommand \enquote  [1]{``#1''}%
\providecommand \bibnamefont  [1]{#1}%
\providecommand \bibfnamefont [1]{#1}%
\providecommand \citenamefont [1]{#1}%
\providecommand \href@noop [0]{\@secondoftwo}%
\providecommand \href [0]{\begingroup \@sanitize@url \@href}%
\providecommand \@href[1]{\@@startlink{#1}\@@href}%
\providecommand \@@href[1]{\endgroup#1\@@endlink}%
\providecommand \@sanitize@url [0]{\catcode `\\12\catcode `\$12\catcode
  `\&12\catcode `\#12\catcode `\^12\catcode `\_12\catcode `\%12\relax}%
\providecommand \@@startlink[1]{}%
\providecommand \@@endlink[0]{}%
\providecommand \url  [0]{\begingroup\@sanitize@url \@url }%
\providecommand \@url [1]{\endgroup\@href {#1}{\urlprefix }}%
\providecommand \urlprefix  [0]{URL }%
\providecommand \Eprint [0]{\href }%
\providecommand \doibase [0]{http://dx.doi.org/}%
\providecommand \selectlanguage [0]{\@gobble}%
\providecommand \bibinfo  [0]{\@secondoftwo}%
\providecommand \bibfield  [0]{\@secondoftwo}%
\providecommand \translation [1]{[#1]}%
\providecommand \BibitemOpen [0]{}%
\providecommand \bibitemStop [0]{}%
\providecommand \bibitemNoStop [0]{.\EOS\space}%
\providecommand \EOS [0]{\spacefactor3000\relax}%
\providecommand \BibitemShut  [1]{\csname bibitem#1\endcsname}%
\let\auto@bib@innerbib\@empty
\bibitem [{\citenamefont {Bakr}\ \emph {et~al.}(2009)\citenamefont {Bakr},
  \citenamefont {Gillen}, \citenamefont {Peng}, \citenamefont {Folling},\ and\
  \citenamefont {Greiner}}]{bakr2009}%
  \BibitemOpen
  \bibfield  {author} {\bibinfo {author} {\bibfnamefont {W.~S.}\ \bibnamefont
  {Bakr}}, \bibinfo {author} {\bibfnamefont {J.~I.}\ \bibnamefont {Gillen}},
  \bibinfo {author} {\bibfnamefont {A.}~\bibnamefont {Peng}}, \bibinfo {author}
  {\bibfnamefont {S.}~\bibnamefont {Folling}}, \ and\ \bibinfo {author}
  {\bibfnamefont {M.}~\bibnamefont {Greiner}},\ }\bibfield  {booktitle} {\emph
  {\bibinfo {booktitle} {Nature}},\ }\href {\doibase 10.1038/nature08482}
  {\bibfield  {journal} {\bibinfo  {journal} {Nature}\ }\textbf {\bibinfo
  {volume} {462}},\ \bibinfo {pages} {74} (\bibinfo {year} {2009})}\BibitemShut
  {NoStop}%
\bibitem [{\citenamefont {Sherson}\ \emph {et~al.}(2010)\citenamefont
  {Sherson}, \citenamefont {Weitenberg}, \citenamefont {Endres}, \citenamefont
  {Cheneau}, \citenamefont {Bloch},\ and\ \citenamefont {Kuhr}}]{sherson2010}%
  \BibitemOpen
  \bibfield  {author} {\bibinfo {author} {\bibfnamefont {J.~F.}\ \bibnamefont
  {Sherson}}, \bibinfo {author} {\bibfnamefont {C.}~\bibnamefont {Weitenberg}},
  \bibinfo {author} {\bibfnamefont {M.}~\bibnamefont {Endres}}, \bibinfo
  {author} {\bibfnamefont {M.}~\bibnamefont {Cheneau}}, \bibinfo {author}
  {\bibfnamefont {I.}~\bibnamefont {Bloch}}, \ and\ \bibinfo {author}
  {\bibfnamefont {S.}~\bibnamefont {Kuhr}},\ }\href {\doibase
  10.1038/nature09378} {\bibfield  {journal} {\bibinfo  {journal} {Nature}\
  }\textbf {\bibinfo {volume} {467}},\ \bibinfo {pages} {68} (\bibinfo {year}
  {2010})}\BibitemShut {NoStop}%
\bibitem [{\citenamefont {Weitenberg}\ \emph {et~al.}(2011)\citenamefont
  {Weitenberg}, \citenamefont {Endres}, \citenamefont {Sherson}, \citenamefont
  {Cheneau}, \citenamefont {Schauss}, \citenamefont {Fukuhara}, \citenamefont
  {Bloch},\ and\ \citenamefont {Kuhr}}]{weitenberg2011single}%
  \BibitemOpen
  \bibfield  {author} {\bibinfo {author} {\bibfnamefont {C.}~\bibnamefont
  {Weitenberg}}, \bibinfo {author} {\bibfnamefont {M.}~\bibnamefont {Endres}},
  \bibinfo {author} {\bibfnamefont {J.~F.}\ \bibnamefont {Sherson}}, \bibinfo
  {author} {\bibfnamefont {M.}~\bibnamefont {Cheneau}}, \bibinfo {author}
  {\bibfnamefont {P.}~\bibnamefont {Schauss}}, \bibinfo {author} {\bibfnamefont
  {T.}~\bibnamefont {Fukuhara}}, \bibinfo {author} {\bibfnamefont
  {I.}~\bibnamefont {Bloch}}, \ and\ \bibinfo {author} {\bibfnamefont
  {S.}~\bibnamefont {Kuhr}},\ }\href {http://dx.doi.org/10.1038/nature09827}
  {\bibfield  {journal} {\bibinfo  {journal} {Nature}\ }\textbf {\bibinfo
  {volume} {471}},\ \bibinfo {pages} {319} (\bibinfo {year}
  {2011})}\BibitemShut {NoStop}%
\bibitem [{\citenamefont {Mandel}\ \emph {et~al.}(2003)\citenamefont {Mandel},
  \citenamefont {Greiner}, \citenamefont {Widera}, \citenamefont {Rom},
  \citenamefont {Hansch},\ and\ \citenamefont {Bloch}}]{mandel2003}%
  \BibitemOpen
  \bibfield  {author} {\bibinfo {author} {\bibfnamefont {O.}~\bibnamefont
  {Mandel}}, \bibinfo {author} {\bibfnamefont {M.}~\bibnamefont {Greiner}},
  \bibinfo {author} {\bibfnamefont {A.}~\bibnamefont {Widera}}, \bibinfo
  {author} {\bibfnamefont {T.}~\bibnamefont {Rom}}, \bibinfo {author}
  {\bibfnamefont {T.~W.}\ \bibnamefont {Hansch}}, \ and\ \bibinfo {author}
  {\bibfnamefont {I.}~\bibnamefont {Bloch}},\ }\href {\doibase
  10.1038/nature02008} {\bibfield  {journal} {\bibinfo  {journal} {Nature}\
  }\textbf {\bibinfo {volume} {425}},\ \bibinfo {pages} {937} (\bibinfo {year}
  {2003})}\BibitemShut {NoStop}%
\bibitem [{\citenamefont {Anderlini}\ \emph {et~al.}(2007)\citenamefont
  {Anderlini}, \citenamefont {Lee}, \citenamefont {Brown}, \citenamefont
  {Sebby-Strabley}, \citenamefont {Phillips},\ and\ \citenamefont
  {Porto}}]{anderlini2007controlled}%
  \BibitemOpen
  \bibfield  {author} {\bibinfo {author} {\bibfnamefont {M.}~\bibnamefont
  {Anderlini}}, \bibinfo {author} {\bibfnamefont {P.~J.}\ \bibnamefont {Lee}},
  \bibinfo {author} {\bibfnamefont {B.~L.}\ \bibnamefont {Brown}}, \bibinfo
  {author} {\bibfnamefont {J.}~\bibnamefont {Sebby-Strabley}}, \bibinfo
  {author} {\bibfnamefont {W.~D.}\ \bibnamefont {Phillips}}, \ and\ \bibinfo
  {author} {\bibfnamefont {J.}~\bibnamefont {Porto}},\ }\href
  {http://dx.doi.org/10.1038/nature06011} {\bibfield  {journal} {\bibinfo
  {journal} {Nature}\ }\textbf {\bibinfo {volume} {448}},\ \bibinfo {pages}
  {452} (\bibinfo {year} {2007})}\BibitemShut {NoStop}%
\bibitem [{\citenamefont {Robens}\ \emph
  {et~al.}(2017{\natexlab{a}})\citenamefont {Robens}, \citenamefont {Zopes},
  \citenamefont {Alt}, \citenamefont {Brakhane}, \citenamefont {Meschede},\
  and\ \citenamefont {Alberti}}]{Robens2017sorting}%
  \BibitemOpen
  \bibfield  {author} {\bibinfo {author} {\bibfnamefont {C.}~\bibnamefont
  {Robens}}, \bibinfo {author} {\bibfnamefont {J.}~\bibnamefont {Zopes}},
  \bibinfo {author} {\bibfnamefont {W.}~\bibnamefont {Alt}}, \bibinfo {author}
  {\bibfnamefont {S.}~\bibnamefont {Brakhane}}, \bibinfo {author}
  {\bibfnamefont {D.}~\bibnamefont {Meschede}}, \ and\ \bibinfo {author}
  {\bibfnamefont {A.}~\bibnamefont {Alberti}},\ }\href {\doibase
  10.1103/PhysRevLett.118.065302} {\bibfield  {journal} {\bibinfo  {journal}
  {Phys. Rev. Lett.}\ }\textbf {\bibinfo {volume} {118}},\ \bibinfo {pages}
  {065302} (\bibinfo {year} {2017}{\natexlab{a}})}\BibitemShut {NoStop}%
\bibitem [{\citenamefont {Gehr}\ \emph {et~al.}(2010)\citenamefont {Gehr},
  \citenamefont {Volz}, \citenamefont {Dubois}, \citenamefont {Steinmetz},
  \citenamefont {Colombe}, \citenamefont {Lev}, \citenamefont {Long},
  \citenamefont {Est\`eve},\ and\ \citenamefont {Reichel}}]{gehr2010}%
  \BibitemOpen
  \bibfield  {author} {\bibinfo {author} {\bibfnamefont {R.}~\bibnamefont
  {Gehr}}, \bibinfo {author} {\bibfnamefont {J.}~\bibnamefont {Volz}}, \bibinfo
  {author} {\bibfnamefont {G.}~\bibnamefont {Dubois}}, \bibinfo {author}
  {\bibfnamefont {T.}~\bibnamefont {Steinmetz}}, \bibinfo {author}
  {\bibfnamefont {Y.}~\bibnamefont {Colombe}}, \bibinfo {author} {\bibfnamefont
  {B.}~\bibnamefont {Lev}}, \bibinfo {author} {\bibfnamefont {R.}~\bibnamefont
  {Long}}, \bibinfo {author} {\bibfnamefont {J.}~\bibnamefont {Est\`eve}}, \
  and\ \bibinfo {author} {\bibfnamefont {J.}~\bibnamefont {Reichel}},\ }\href
  {\doibase 10.1103/PhysRevLett.104.203602} {\bibfield  {journal} {\bibinfo
  {journal} {Phys. Rev. Lett.}\ }\textbf {\bibinfo {volume} {104}},\ \bibinfo
  {pages} {203602} (\bibinfo {year} {2010})}\BibitemShut {NoStop}%
\bibitem [{\citenamefont {Bochmann}\ \emph {et~al.}(2010)\citenamefont
  {Bochmann}, \citenamefont {M\"ucke}, \citenamefont {Guhl}, \citenamefont
  {Ritter}, \citenamefont {Rempe},\ and\ \citenamefont
  {Moehring}}]{bochmann2010}%
  \BibitemOpen
  \bibfield  {author} {\bibinfo {author} {\bibfnamefont {J.}~\bibnamefont
  {Bochmann}}, \bibinfo {author} {\bibfnamefont {M.}~\bibnamefont {M\"ucke}},
  \bibinfo {author} {\bibfnamefont {C.}~\bibnamefont {Guhl}}, \bibinfo {author}
  {\bibfnamefont {S.}~\bibnamefont {Ritter}}, \bibinfo {author} {\bibfnamefont
  {G.}~\bibnamefont {Rempe}}, \ and\ \bibinfo {author} {\bibfnamefont
  {D.}~\bibnamefont {Moehring}},\ }\href {\doibase
  10.1103/PhysRevLett.104.203601} {\bibfield  {journal} {\bibinfo  {journal}
  {Phys. Rev. Lett.}\ }\textbf {\bibinfo {volume} {104}},\ \bibinfo {pages}
  {203601} (\bibinfo {year} {2010})}\BibitemShut {NoStop}%
\bibitem [{\citenamefont {Reick}\ \emph {et~al.}(2010)\citenamefont {Reick},
  \citenamefont {M{\o}lmer}, \citenamefont {Alt}, \citenamefont {Eckstein},
  \citenamefont {Kampschulte}, \citenamefont {Kong}, \citenamefont {Reimann},
  \citenamefont {Thobe}, \citenamefont {Widera},\ and\ \citenamefont
  {Meschede}}]{reick2010}%
  \BibitemOpen
  \bibfield  {author} {\bibinfo {author} {\bibfnamefont {S.}~\bibnamefont
  {Reick}}, \bibinfo {author} {\bibfnamefont {K.}~\bibnamefont {M{\o}lmer}},
  \bibinfo {author} {\bibfnamefont {W.}~\bibnamefont {Alt}}, \bibinfo {author}
  {\bibfnamefont {M.}~\bibnamefont {Eckstein}}, \bibinfo {author}
  {\bibfnamefont {T.}~\bibnamefont {Kampschulte}}, \bibinfo {author}
  {\bibfnamefont {L.}~\bibnamefont {Kong}}, \bibinfo {author} {\bibfnamefont
  {R.}~\bibnamefont {Reimann}}, \bibinfo {author} {\bibfnamefont
  {A.}~\bibnamefont {Thobe}}, \bibinfo {author} {\bibfnamefont
  {A.}~\bibnamefont {Widera}}, \ and\ \bibinfo {author} {\bibfnamefont
  {D.}~\bibnamefont {Meschede}},\ }\href {\doibase 10.1364/josab.27.00a152}
  {\bibfield  {journal} {\bibinfo  {journal} {J. Opt. Soc. Am. B}\ }\textbf
  {\bibinfo {volume} {27}},\ \bibinfo {pages} {A152} (\bibinfo {year}
  {2010})}\BibitemShut {NoStop}%
\bibitem [{\citenamefont {Gibbons}\ \emph {et~al.}(2011)\citenamefont
  {Gibbons}, \citenamefont {Hamley}, \citenamefont {Shih},\ and\ \citenamefont
  {Chapman}}]{gibbons2011}%
  \BibitemOpen
  \bibfield  {author} {\bibinfo {author} {\bibfnamefont {M.~J.}\ \bibnamefont
  {Gibbons}}, \bibinfo {author} {\bibfnamefont {C.~D.}\ \bibnamefont {Hamley}},
  \bibinfo {author} {\bibfnamefont {C.-Y.}\ \bibnamefont {Shih}}, \ and\
  \bibinfo {author} {\bibfnamefont {M.~S.}\ \bibnamefont {Chapman}},\ }\href
  {\doibase 10.1103/PhysRevLett.106.133002} {\bibfield  {journal} {\bibinfo
  {journal} {Phys. Rev. Lett.}\ }\textbf {\bibinfo {volume} {106}},\ \bibinfo
  {pages} {133002} (\bibinfo {year} {2011})}\BibitemShut {NoStop}%
\bibitem [{\citenamefont {Fuhrmanek}\ \emph {et~al.}(2011)\citenamefont
  {Fuhrmanek}, \citenamefont {Bourgain}, \citenamefont {Sortais},\ and\
  \citenamefont {Browaeys}}]{fuhrmanek2011}%
  \BibitemOpen
  \bibfield  {author} {\bibinfo {author} {\bibfnamefont {A.}~\bibnamefont
  {Fuhrmanek}}, \bibinfo {author} {\bibfnamefont {R.}~\bibnamefont {Bourgain}},
  \bibinfo {author} {\bibfnamefont {Y.~R.~P.}\ \bibnamefont {Sortais}}, \ and\
  \bibinfo {author} {\bibfnamefont {A.}~\bibnamefont {Browaeys}},\ }\href
  {\doibase 10.1103/PhysRevLett.106.133003} {\bibfield  {journal} {\bibinfo
  {journal} {Phys. Rev. Lett.}\ }\textbf {\bibinfo {volume} {106}},\ \bibinfo
  {pages} {133003} (\bibinfo {year} {2011})}\BibitemShut {NoStop}%
\bibitem [{\citenamefont {Robens}\ \emph
  {et~al.}(2017{\natexlab{b}})\citenamefont {Robens}, \citenamefont {Alt},
  \citenamefont {Emary}, \citenamefont {Meschede},\ and\ \citenamefont
  {Alberti}}]{robens2017atomic}%
  \BibitemOpen
  \bibfield  {author} {\bibinfo {author} {\bibfnamefont {C.}~\bibnamefont
  {Robens}}, \bibinfo {author} {\bibfnamefont {W.}~\bibnamefont {Alt}},
  \bibinfo {author} {\bibfnamefont {C.}~\bibnamefont {Emary}}, \bibinfo
  {author} {\bibfnamefont {D.}~\bibnamefont {Meschede}}, \ and\ \bibinfo
  {author} {\bibfnamefont {A.}~\bibnamefont {Alberti}},\ }\href
  {https://link.springer.com/article/10.1007%2Fs00340-016-6581-y} {\bibfield
  {journal} {\bibinfo  {journal} {Applied Physics B}\ }\textbf {\bibinfo
  {volume} {123}},\ \bibinfo {pages} {12} (\bibinfo {year}
  {2017}{\natexlab{b}})}\BibitemShut {NoStop}%
\bibitem [{\citenamefont {Boll}\ \emph {et~al.}(2016)\citenamefont {Boll},
  \citenamefont {Hilker}, \citenamefont {Salomon}, \citenamefont {Omran},
  \citenamefont {Nespolo}, \citenamefont {Pollet}, \citenamefont {Bloch},\ and\
  \citenamefont {Gross}}]{boll2016spin}%
  \BibitemOpen
  \bibfield  {author} {\bibinfo {author} {\bibfnamefont {M.}~\bibnamefont
  {Boll}}, \bibinfo {author} {\bibfnamefont {T.~A.}\ \bibnamefont {Hilker}},
  \bibinfo {author} {\bibfnamefont {G.}~\bibnamefont {Salomon}}, \bibinfo
  {author} {\bibfnamefont {A.}~\bibnamefont {Omran}}, \bibinfo {author}
  {\bibfnamefont {J.}~\bibnamefont {Nespolo}}, \bibinfo {author} {\bibfnamefont
  {L.}~\bibnamefont {Pollet}}, \bibinfo {author} {\bibfnamefont
  {I.}~\bibnamefont {Bloch}}, \ and\ \bibinfo {author} {\bibfnamefont
  {C.}~\bibnamefont {Gross}},\ }\href {\doibase 10.1126/science.aag1635}
  {\bibfield  {journal} {\bibinfo  {journal} {Science}\ }\textbf {\bibinfo
  {volume} {353}},\ \bibinfo {pages} {1257} (\bibinfo {year}
  {2016})}\BibitemShut {NoStop}%
\bibitem [{\citenamefont {Meyer}(1996)}]{Meyer1996}%
  \BibitemOpen
  \bibfield  {author} {\bibinfo {author} {\bibfnamefont {D.~A.}\ \bibnamefont
  {Meyer}},\ }\href {\doibase 10.1007/BF02199356} {\bibfield  {journal}
  {\bibinfo  {journal} {Journal of Statistical Physics}\ }\textbf {\bibinfo
  {volume} {85}},\ \bibinfo {pages} {551} (\bibinfo {year} {1996})}\BibitemShut
  {NoStop}%
\bibitem [{\citenamefont {Karski}\ \emph {et~al.}(2009)\citenamefont {Karski},
  \citenamefont {F{\"o}rster}, \citenamefont {Choi}, \citenamefont {Steffen},
  \citenamefont {Alt}, \citenamefont {Meschede},\ and\ \citenamefont
  {Widera}}]{karski2009quantum}%
  \BibitemOpen
  \bibfield  {author} {\bibinfo {author} {\bibfnamefont {M.}~\bibnamefont
  {Karski}}, \bibinfo {author} {\bibfnamefont {L.}~\bibnamefont {F{\"o}rster}},
  \bibinfo {author} {\bibfnamefont {J.-M.}\ \bibnamefont {Choi}}, \bibinfo
  {author} {\bibfnamefont {A.}~\bibnamefont {Steffen}}, \bibinfo {author}
  {\bibfnamefont {W.}~\bibnamefont {Alt}}, \bibinfo {author} {\bibfnamefont
  {D.}~\bibnamefont {Meschede}}, \ and\ \bibinfo {author} {\bibfnamefont
  {A.}~\bibnamefont {Widera}},\ }\href {\doibase 10.1126/science.1174436}
  {\bibfield  {journal} {\bibinfo  {journal} {Science}\ }\textbf {\bibinfo
  {volume} {325}},\ \bibinfo {pages} {174} (\bibinfo {year}
  {2009})}\BibitemShut {NoStop}%
\bibitem [{\citenamefont {Alberti}\ \emph {et~al.}(2016)\citenamefont
  {Alberti}, \citenamefont {Robens}, \citenamefont {Alt}, \citenamefont
  {Brakhane}, \citenamefont {Karski}, \citenamefont {Reimann}, \citenamefont
  {Widera},\ and\ \citenamefont {Meschede}}]{Alberti2016SuperResolution}%
  \BibitemOpen
  \bibfield  {author} {\bibinfo {author} {\bibfnamefont {A.}~\bibnamefont
  {Alberti}}, \bibinfo {author} {\bibfnamefont {C.}~\bibnamefont {Robens}},
  \bibinfo {author} {\bibfnamefont {W.}~\bibnamefont {Alt}}, \bibinfo {author}
  {\bibfnamefont {S.}~\bibnamefont {Brakhane}}, \bibinfo {author}
  {\bibfnamefont {M.}~\bibnamefont {Karski}}, \bibinfo {author} {\bibfnamefont
  {R.}~\bibnamefont {Reimann}}, \bibinfo {author} {\bibfnamefont
  {A.}~\bibnamefont {Widera}}, \ and\ \bibinfo {author} {\bibfnamefont
  {D.}~\bibnamefont {Meschede}},\ }\href
  {http://stacks.iop.org/1367-2630/18/i=5/a=053010} {\bibfield  {journal}
  {\bibinfo  {journal} {New Journal of Physics}\ }\textbf {\bibinfo {volume}
  {18}},\ \bibinfo {pages} {053010} (\bibinfo {year} {2016})}\BibitemShut
  {NoStop}%
\bibitem [{Note1()}]{Note1}%
  \BibitemOpen
  \bibinfo {note} {Whereas we have implemented $m_F$-pumping to initialize the
  bright state in the experiment, we only use $F$-state pumping for the
  preparation of the dark states, since all Zeeman sub-levels of the $F=1$
  state are equally dark.}\BibitemShut {Stop}%
\bibitem [{\citenamefont {Dalibard}\ and\ \citenamefont
  {Cohen-Tannoudji}(1985)}]{dalibard1985}%
  \BibitemOpen
  \bibfield  {author} {\bibinfo {author} {\bibfnamefont {J.}~\bibnamefont
  {Dalibard}}\ and\ \bibinfo {author} {\bibfnamefont {C.}~\bibnamefont
  {Cohen-Tannoudji}},\ }\href {\doibase 10.1364/JOSAB.2.001707} {\bibfield
  {journal} {\bibinfo  {journal} {J. Opt. Soc. Am. B}\ }\textbf {\bibinfo
  {volume} {2}},\ \bibinfo {pages} {1707} (\bibinfo {year} {1985})}\BibitemShut
  {NoStop}%
\bibitem [{\citenamefont {Cheuk}\ \emph {et~al.}(2015)\citenamefont {Cheuk},
  \citenamefont {Nichols}, \citenamefont {Okan}, \citenamefont {Gersdorf},
  \citenamefont {Ramasesh}, \citenamefont {Bakr}, \citenamefont {Lompe},\ and\
  \citenamefont {Zwierlein}}]{cheuk2015quantum}%
  \BibitemOpen
  \bibfield  {author} {\bibinfo {author} {\bibfnamefont {L.~W.}\ \bibnamefont
  {Cheuk}}, \bibinfo {author} {\bibfnamefont {M.~A.}\ \bibnamefont {Nichols}},
  \bibinfo {author} {\bibfnamefont {M.}~\bibnamefont {Okan}}, \bibinfo {author}
  {\bibfnamefont {T.}~\bibnamefont {Gersdorf}}, \bibinfo {author}
  {\bibfnamefont {V.~V.}\ \bibnamefont {Ramasesh}}, \bibinfo {author}
  {\bibfnamefont {W.~S.}\ \bibnamefont {Bakr}}, \bibinfo {author}
  {\bibfnamefont {T.}~\bibnamefont {Lompe}}, \ and\ \bibinfo {author}
  {\bibfnamefont {M.~W.}\ \bibnamefont {Zwierlein}},\ }\href {\doibase
  10.1103/PhysRevLett.114.193001} {\bibfield  {journal} {\bibinfo  {journal}
  {Phys. Rev. Lett.}\ }\textbf {\bibinfo {volume} {114}},\ \bibinfo {pages}
  {193001} (\bibinfo {year} {2015})}\BibitemShut {NoStop}%
\bibitem [{Note2()}]{Note2}%
  \BibitemOpen
  \bibinfo {note} {We have used a harmonic approximation for the ground state
  potential and assumed that the excited state lifetime $1/\Gamma $ is much
  shorter than the trap oscillation period}\BibitemShut {NoStop}%
\bibitem [{\citenamefont {Martinez-Dorantes}\ \emph {et~al.}()\citenamefont
  {Martinez-Dorantes}, \citenamefont {Alt}, \citenamefont {Gallego},
  \citenamefont {Ghosh}, \citenamefont {Ratschbacher},\ and\ \citenamefont
  {Meschede}}]{LongPaper}%
  \BibitemOpen
  \bibfield  {author} {\bibinfo {author} {\bibfnamefont {M.}~\bibnamefont
  {Martinez-Dorantes}}, \bibinfo {author} {\bibfnamefont {W.}~\bibnamefont
  {Alt}}, \bibinfo {author} {\bibfnamefont {J.}~\bibnamefont {Gallego}},
  \bibinfo {author} {\bibfnamefont {S.}~\bibnamefont {Ghosh}}, \bibinfo
  {author} {\bibfnamefont {L.}~\bibnamefont {Ratschbacher}}, \ and\ \bibinfo
  {author} {\bibfnamefont {D.}~\bibnamefont {Meschede}},\ }\href@noop {}
  {\bibinfo  {journal} {To be published}\ }\BibitemShut {NoStop}%
\bibitem [{\citenamefont {Sivia}\ and\ \citenamefont
  {Skilling}(2006)}]{sivia2006data}%
  \BibitemOpen
\bibfield  {journal} {  }\bibfield  {author} {\bibinfo {author} {\bibfnamefont
  {D.}~\bibnamefont {Sivia}}\ and\ \bibinfo {author} {\bibfnamefont
  {J.}~\bibnamefont {Skilling}},\ }\href@noop {} {\emph {\bibinfo {title} {Data
  analysis: a Bayesian tutorial}}}\ (\bibinfo  {publisher} {OUP Oxford},\
  \bibinfo {year} {2006})\BibitemShut {NoStop}%
\bibitem [{Note3()}]{Note3}%
  \BibitemOpen
  \bibinfo {note} {Using the full two-dimensional intensity distribution, i.e.
  pixels instead of columns, does not noticeably improve state detection but
  incurs a drastic increase of computational effort.}\BibitemShut {Stop}%
\bibitem [{\citenamefont {Acton}\ \emph {et~al.}(2006)\citenamefont {Acton},
  \citenamefont {Brickman}, \citenamefont {Haljan}, \citenamefont {Lee},
  \citenamefont {Deslauriers},\ and\ \citenamefont {Monroe}}]{acton2006}%
  \BibitemOpen
  \bibfield  {author} {\bibinfo {author} {\bibfnamefont {M.}~\bibnamefont
  {Acton}}, \bibinfo {author} {\bibfnamefont {K.-A.}\ \bibnamefont {Brickman}},
  \bibinfo {author} {\bibfnamefont {P.~C.}\ \bibnamefont {Haljan}}, \bibinfo
  {author} {\bibfnamefont {P.~J.}\ \bibnamefont {Lee}}, \bibinfo {author}
  {\bibfnamefont {L.}~\bibnamefont {Deslauriers}}, \ and\ \bibinfo {author}
  {\bibfnamefont {C.}~\bibnamefont {Monroe}},\ }\href
  {http://dl.acm.org/citation.cfm?id=2011691.2011692} {\bibfield  {journal}
  {\bibinfo  {journal} {Quantum Info. Comput.}\ }\textbf {\bibinfo {volume}
  {6}},\ \bibinfo {pages} {465} (\bibinfo {year} {2006})}\BibitemShut {NoStop}%
\bibitem [{\citenamefont {Robbins}(2011)}]{Robbins2011}%
  \BibitemOpen
  \bibfield  {author} {\bibinfo {author} {\bibfnamefont {M.~S.}\ \bibnamefont
  {Robbins}},\ }\enquote {\bibinfo {title} {Electron-multiplying charge coupled
  devices -- emccds},}\ in\ \href {\doibase 10.1007/978-3-642-18443-7_6} {\emph
  {\bibinfo {booktitle} {Single-Photon Imaging}}},\ \bibinfo {editor} {edited
  by\ \bibinfo {editor} {\bibfnamefont {P.}~\bibnamefont {Seitz}}\ and\
  \bibinfo {editor} {\bibfnamefont {A.~J.}\ \bibnamefont {Theuwissen}}}\
  (\bibinfo  {publisher} {Springer Berlin Heidelberg},\ \bibinfo {address}
  {Berlin, Heidelberg},\ \bibinfo {year} {2011})\ pp.\ \bibinfo {pages}
  {103--121}\BibitemShut {NoStop}%
\bibitem [{\citenamefont {Harps{\o}e}\ \emph {et~al.}(2012)\citenamefont
  {Harps{\o}e}, \citenamefont {Andersen},\ and\ \citenamefont
  {Kj{\ae}gaard}}]{citeulike:10216617}%
  \BibitemOpen
  \bibfield  {author} {\bibinfo {author} {\bibfnamefont {K.~B.~W.}\
  \bibnamefont {Harps{\o}e}}, \bibinfo {author} {\bibfnamefont {M.~I.}\
  \bibnamefont {Andersen}}, \ and\ \bibinfo {author} {\bibfnamefont
  {P.}~\bibnamefont {Kj{\ae}gaard}},\ }\href {\doibase
  10.1051/0004-6361/201117089} {\bibfield  {journal} {\bibinfo  {journal}
  {Astronomy \& Astrophysics}\ }\textbf {\bibinfo {volume} {537}},\ \bibinfo
  {pages} {A50+} (\bibinfo {year} {2012})}\BibitemShut {NoStop}%
\bibitem [{\citenamefont {Parsons}\ \emph {et~al.}(2015)\citenamefont
  {Parsons}, \citenamefont {Huber}, \citenamefont {Mazurenko}, \citenamefont
  {Chiu}, \citenamefont {Setiawan}, \citenamefont {Wooley-Brown}, \citenamefont
  {Blatt},\ and\ \citenamefont {Greiner}}]{parsons2015site}%
  \BibitemOpen
  \bibfield  {author} {\bibinfo {author} {\bibfnamefont {M.~F.}\ \bibnamefont
  {Parsons}}, \bibinfo {author} {\bibfnamefont {F.}~\bibnamefont {Huber}},
  \bibinfo {author} {\bibfnamefont {A.}~\bibnamefont {Mazurenko}}, \bibinfo
  {author} {\bibfnamefont {C.~S.}\ \bibnamefont {Chiu}}, \bibinfo {author}
  {\bibfnamefont {W.}~\bibnamefont {Setiawan}}, \bibinfo {author}
  {\bibfnamefont {K.}~\bibnamefont {Wooley-Brown}}, \bibinfo {author}
  {\bibfnamefont {S.}~\bibnamefont {Blatt}}, \ and\ \bibinfo {author}
  {\bibfnamefont {M.}~\bibnamefont {Greiner}},\ }\href {\doibase
  10.1103/PhysRevLett.114.213002} {\bibfield  {journal} {\bibinfo  {journal}
  {Phys. Rev. Lett.}\ }\textbf {\bibinfo {volume} {114}},\ \bibinfo {pages}
  {213002} (\bibinfo {year} {2015})}\BibitemShut {NoStop}%
\bibitem [{\citenamefont {Burrell}\ \emph {et~al.}(2010)\citenamefont
  {Burrell}, \citenamefont {Szwer}, \citenamefont {Webster},\ and\
  \citenamefont {Lucas}}]{burrell2010}%
  \BibitemOpen
  \bibfield  {author} {\bibinfo {author} {\bibfnamefont {A.~H.}\ \bibnamefont
  {Burrell}}, \bibinfo {author} {\bibfnamefont {D.~J.}\ \bibnamefont {Szwer}},
  \bibinfo {author} {\bibfnamefont {S.~C.}\ \bibnamefont {Webster}}, \ and\
  \bibinfo {author} {\bibfnamefont {D.~M.}\ \bibnamefont {Lucas}},\ }\href
  {\doibase 10.1103/PhysRevA.81.040302} {\bibfield  {journal} {\bibinfo
  {journal} {Phys. Rev. A}\ }\textbf {\bibinfo {volume} {81}},\ \bibinfo
  {pages} {040302} (\bibinfo {year} {2010})}\BibitemShut {NoStop}%
\end{thebibliography}%
\bibliographystyle{apsrev4-1}
\end{document}